\documentclass[nofootinbib,reprint,amsmath,amssymb,showkeys, showpacs,aps]{revtex4-1}
\usepackage{graphicx}
\usepackage[T1]{fontenc}
\usepackage{dcolumn}
\usepackage{xcolor,colortbl}
\usepackage{multirow}
\usepackage{bm}
\newcommand{\qm}[1]{``#1''}
\usepackage{hyperref}
\hypersetup{colorlinks, linkcolor={red},citecolor={blue},urlcolor={blue}}  
\newcommand\ChangeRT[1]{\noalign{\hrule height #1}}

\begin{document}

\preprint{APS/123-QED}

\title[Epicyclic frequencies in static and spherically symmetric wormhole geometries]{Epicyclic frequencies in static and spherically symmetric wormhole geometries}

\author{Vittorio De Falco$^{1,2}$}\email{vittorio.defalco@physics.cz}
\author{Mariafelicia De Laurentis$^{3,2,4}$\vspace{0.5cm}}\email{mariafelicia.delaurentis@unina.it}
\author{Salvatore Capozziello$^{2,3,4,5}$}\email{capozziello@unina.it}

\affiliation{$^1$Department of Mathematics and Applications \qm{R. Caccioppoli}, University of Naples Federico II, Via Cintia, 80126 Naples, Italy,\\
$^2$ Istituto Nazionale di Fisica Nucleare, Sezione di Napoli, Complesso Universitario di Monte S. Angelo, Via Cintia Edificio 6, 80126 Napoli, Italy\\
$^3$ Universit\`{a} degli studi di Napoli \qm{Federico II}, Dipartimento di Fisica \qm{Ettore Pancini}, Complesso Universitario di Monte S. Angelo, Via Cintia Edificio 6, 80126 Napoli, Italy\\
$^4$ Lab.Theor.Cosmology,Tomsk State University of Control Systems and Radioelectronics(TUSUR), 634050 Tomsk, Russia\\
$^5$Scuola Superiore Meridionale, Universit\`{a}  di Napoli \qm{Federico II},  Largo San Marcellino 10, 80138 Napoli, Italy}

\date{\today}

\begin{abstract}
The measurement of the epicyclic frequencies is a widely used astrophysical technique to infer information on a given self-gravitating system and on the related gravity background. We derive their explicit expressions in static and spherically symmetric wormhole spacetimes. We discuss how these theoretical results can be applied to: (1) detect the presence of a wormhole, distinguishing it by a black hole; (2) reconstruct wormhole solutions through the fit of the observational data, once we have them. Finally, we discuss the physical implications of our proposed epicyclic method.   
\end{abstract}
\pacs{04.20.Dw, 04.70as, 04.25.dg}
\keywords{Physics of black holes, alternative gravity, wormhole.}

\maketitle
\section{Introduction}
\label{sec:intro}
A wormhole (WH) can be intuitively seen as a topological shortcut-structure, which capable of  connecting two distinct spacetime points. A visual representation of a WH is obtained through the example of drawing two separate points on a paper sheet, and then considering as the shortest connecting-trajectory not the joining straight line, but the bent paper which brings the two points one over the other. 
\begin{figure}[h!]
\centering
\includegraphics[scale=0.32]{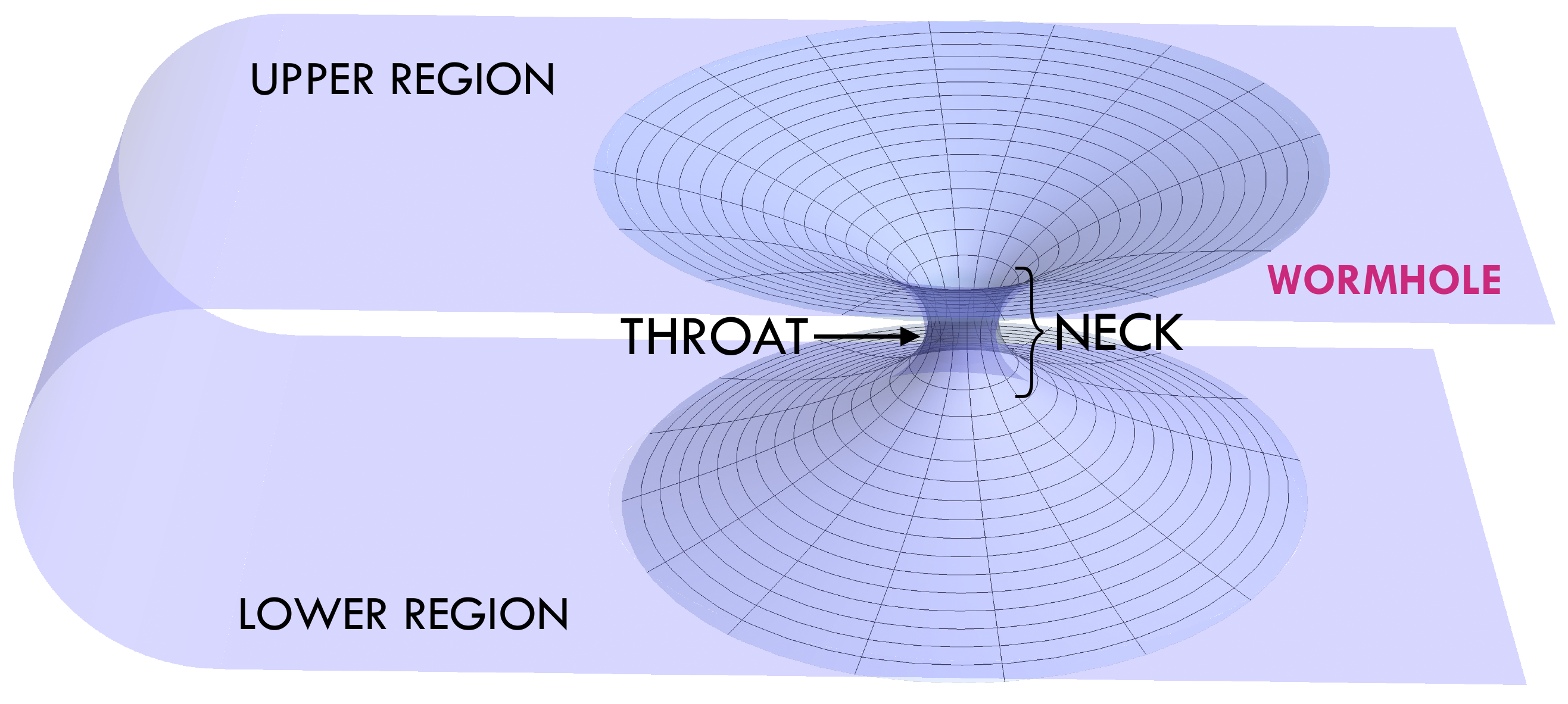}
\caption{Geometrical representation of a WH.}
\label{fig:Fig0}
\end{figure}
In Fig. \ref{fig:Fig0}, we sketch   the intuitive picture reported before. A WH is conceptually  defined as a compact object characterized by no horizons and physical singularities, and endowed with a traversable bridge, dubbed WH neck, connecting two universes or two different regions of the same spacetime \cite{Visser1995}. 

These objects have been extensively studied in the literature, indeed there are many authors, who not only built up new WH solutions, both in General Relativity (GR) and in Alternative Gravity, but they were also interested in analysing their properties \cite{Visser1989,Barcelo1999,Bohmer2012,Anchordoqui:2000ut,Bahamonde:2016jqq,Capozziello2020,Lorenza}. On the other hand, there is also a major research  effort in conceiving original astrophysical strategies to search for WH observational signatures \cite{Li2014,Cardoso2016,Konoplya2016,Paul2019,Dai2019,Banerjee2019,Hashimoto2017,Dalui2019,Defalco2020WH,DeFalco2021,DeFalco2021WF}. This research topics is strongly motivated not only by the presence of several complementary data, but also because there is the great opportunity to perform, now and in the near-future, highly-precise observations in strong field regimes. This is a very crucial point, because the missed detection of WHs can be so far explained by the fact that gravity has not been investigated in extreme regimes. This justifies also the possibility to find a particular subclass of WH solutions, known in the literature as \emph{black hole (BH) mimickers}, which perfectly mimic all observational properties of a BH with arbitrary accuracy \cite{Cardoso2019} and their signature may be likely revealed by strong gravity experiments.

In this respect, an important role can be played by the \emph{epicyclic frequencies}. Let us assume that a particle moving in a closed orbit is disturbed by small perturbations in the radial, azimuthal, and polar directions. The particle oscillation frequencies, along the above-mentioned directions, correspond to the epicyclic frequencies $\left\{\nu_r,\nu_\varphi,\nu_\theta\right\}$, respectively. These quantities reveal some useful features \cite{Motta2016,Ingram2020}: (1) strong dependence on the underlying spacetime geometry; (2) production of observational effects in strong field regime; (3) direct possibility in measuring them with actual and near future data. At the best of our knowledge, there are only two papers on this subject applied to WHs, which are: (1) Chakraborty and collaborators, who studied the behaviour of a test gyroscope moving towards a Teo rotating traversable WH \cite{Chakraborty2017}; (2) Deligianni and coauthors, who focused their attention on 
quasi-periodic oscillations (QPOs) from an accretion disk around Teo rotating traversable WHs  \cite{Deligianni2021}. 

In this work, we adopt the above-cited advantages of the epicyclic frequencies to elaborate an astrophysical procedure to both observationally unearth WHs and to identify the most appropriate WH solution/s to fit the observational data. Our analysis concentrates on static and spherically symmetric WH metrics. The article is organized as follows: in Sec. \ref{sec:WHEPI}, we summarize our model-independent approach framed in generic static and spherically symmetric WH geometries and then derive the formulas of the epicyclic frequencies; in Sec. \ref{sec:app} we apply the approach  to both observationally detect the presence of a WH and to reconstruct the WH solutions through the fit of the observational data; finally, in Sec. \ref{sec:end}, we discuss the obtained results and draw the conclusions. 

\section{Wormhole epicyclic frequencies}
\label{sec:WHEPI}
In this section, we first describe the general theory in which the WH solutions are framed (see Sec. \ref{sec:WH}). As second step, we consider the timelike geodesic equations (see Sec. \ref{sec:TLG}), and then finally derive the general expressions of the epicyclic frequencies (see Sec. \ref{sec:EF}).

\subsection{The Morris-Thorne-like wormhole metrics}
\label{sec:WH}
A generic static and spherically symmetric WH can be described by the Morris-Thorne-like metric \cite{Morris1988}, $ds^2=g_{\alpha\beta}dx^\alpha dx^\beta$, which in geometrical units ($G=c=1$), in spherical coordinates $(t,r,\theta,\varphi)$, and set in the equatorial plane $\theta=\pi/2$ (without loss of generality due to the spherical symmetry hypothesis) reads as
\begin{equation}  \label{eq:MTmetric}
ds^2=-e^{2\Phi(r)}dt^2+\frac{dr^2}{1-b(r)/r}+r^2d\varphi^2,
\end{equation}
where $\Phi(r)$ and $b(r)$ are the redshift and shape functions, respectively. Eq. (\ref{eq:MTmetric}) is a two-parameters family of metrics, fully determined once $\Phi(r)$ and $b(r)$ are known, and it represents also a class of solutions valid both in GR and in several Alternative  Gravity theories, with the further request to be traversable and stable. $\Phi(r)$ and $b(r)$ can be assigned in two  ways: (1) \emph{a-priori}, meaning that a new theoretical WH solution has been found; (2) \emph{a-posteriori}, referring to the fact that they can be reconstructed through the fit of the observational data (approach followed in this paper).

We require that these WH solutions satisfy the following properties \cite{Morris1988}: (1) $\Phi(r)$ and $b(r)$ are real smooth functions, and $\Phi(r)$ is everywhere finite, since there is the 
absence of horizons and essential singularities; to define a finite proper radial distance $l$, we have that $(1-b(r)/r)\ge 0$; (3) the flaring outward condition \cite{Hochberg19981,Hochberg19982,caplobo1,caplobo2} requires that $b^{\prime}(r) < b(r)/r$ near and on the throat. It defines the minimum radius such that $r_{\rm min}=b_0$ and $b(r_{\rm min})=b_0$; (4) asymptotic flatness, namely $b(r)/r\to0$ and $\Phi(r)\to0$ for $r\to + \infty$; (5) the WH traversability, which depending on the gravity framework, can be obtained by considering exotic matter (especially in GR) \cite{Hochberg1997,Bronnikov2013,Garattini2019} or topological defects (mainly in alternative and extended  theories of gravity) \cite{Lobo2009,Harko2013,Capozziello2012}; (6) the mass $M$ is defined, according to the Arnowitt, Deser, Misner (ADM) formalism, as the total mass of the system contained in the whole spacetime \cite{Visser1995}: 
\begin{equation} \label{eq:ADMmass}
M\equiv \lim_{r\to+\infty}m(r)=\frac{c^2 b_0}{2G}+4\pi c^2\int^{\infty}_{b_0}\rho(x)x^2 dx.
\end{equation}

\subsection{Timelike geodesic equations}
\label{sec:TLG}
A test particle moving around any self-gravitating object, in particular a WH, and affected only by gravity follows a timelike geodesic equation, which is
\begin{equation} \label{eq:GE}
\frac{\rm d^2 x^\alpha}{\rm d\tau^2}+\Gamma^\alpha_{\beta\gamma}\frac{\rm d x^\beta}{\rm d\tau}\frac{\rm d x^\gamma}{\rm d\tau}=0,
\end{equation}
where $\tau$  is the affine parameter (proper time) along the test particle trajectory, and $\Gamma^\alpha_{\beta\gamma}$ are the Christoffel symbols. We prefer to adopt the \emph{relativity of observer splitting formalism}, which permits to clearly distinguish between gravitational and inertial contributions (see \cite{Jantzen1992,Bini1997a,Bini1997b,Bini1998,Bini1999,Defalco2018}, for more details). This approach is equivalent to Eq. (\ref{eq:GE}), but it has the great advantage that has a direct connection with the classical description, allowing to understand the physics behind the symbols we algebraically manipulate. We are also aware that there are more direct formulas to calculate the epicyclic frequencies (see Refs. \cite{Chakraborty2017,Deligianni2021}, for details), but we deem however pedagogical to present the full derivation.

The approach can be formulated considering the  presence of two observers: (1) one static located at infinity, corresponding to our telescopes and detectors with which we normally perform observations, and measurements in astrophysics; (2) \emph{local static observers} (LSOs), in which it is more easy performing the calculations. A proper reference frame adapted to the LSOs is given by the orthonormal basis of vectors 
 \cite{Morris1988,Defalco2020WH}
\begin{equation} \label{eq:SOframe}
\begin{aligned}
&\boldsymbol{e_{\hat t}}= \frac{\boldsymbol{\partial_t}}{e^{\Phi(r)}},\quad
\boldsymbol{e_{\hat r}}=\boldsymbol{\partial_r}\sqrt{1-\frac{b(r)}{r}},\quad
\boldsymbol{e_{\hat \varphi}}=\frac{\boldsymbol{\partial_\varphi}}{r}.
\end{aligned}
\end{equation}
We will denote throughout the paper, vector, and tensor indices (e.g., $v_\alpha$; $T_{\alpha\beta}$) evaluated in the LSO frame by a hat (e.g., $v_{\hat \alpha}$; $T_{\hat{\alpha}\hat{\beta}}$), whereas scalar quantities (e.g., $f$) are followed by $n$ (e.g., $f(n)$). A test particle moves with a timelike four-velocity $\boldsymbol{U}$ and a spatial velocity $\boldsymbol{\nu}(U,n)$ with respect to the LSO frames, which both read as \cite{Defalco2020WH}
\begin{equation} \label{eq:NU}
\boldsymbol{U}=\gamma[\boldsymbol{e_{\hat t}}+\boldsymbol{\nu}],\qquad \boldsymbol{\nu}=\nu(\sin\alpha\boldsymbol{e_{\hat r}}+\cos\alpha \boldsymbol{e_{\hat\varphi}}),
\end{equation}
where $\gamma=1/\sqrt{1-\nu^2}$ is the Lorentz factor, $\nu=||\boldsymbol{\nu}||$ is the magnitude of the test particle spatial velocity, and $\alpha$ is the azimuthal angle of the vector $\boldsymbol{\nu}$ measured clockwise from the positive $\boldsymbol{\hat\varphi}$ direction in the LSO frame. 

An important role is played by the \emph{LSO kinematical quantities}, which are: the \emph{acceleration} $a(n)^{\hat r}$, being the general relativistic gravitational attraction along the radial direction, and  the \emph{relative Lie curvature vector} $k_{\rm (Lie)}(n)^{\hat r}$, corresponding to the general relativistic centrifugal force along the radial direction. Their explicit expressions are \cite{Defalco2020WH}
\begin{eqnarray}
a(n)^{\hat r}&=&\Phi'(r) \sqrt{1-\frac{b(r)}{r}},\label{eq:acc}\\
k_{\rm (Lie)}(n)^{\hat r}&=&-\frac{1}{r}\sqrt{1-\frac{b(r)}{r}} \label{eq:klie}.
\end{eqnarray}
where $'=d/dr$. The components of the test particle acceleration $\boldsymbol{a}(U)$ can be calculated as \cite{Bini1997a,Bini1997b,Defalco2018,Defalco2020WH}
\begin{eqnarray}
a(U)^{\hat t}&=&\gamma^2\nu\sin\alpha\ a(n)^{\hat r}+\gamma^3\nu\frac{d\nu}{d\tau}, \label{eq:EOM1}\\
a(U)^{\hat r}&=&\gamma^2[a(n)^{\hat r}+k_{\rm (Lie)}(n)^{\hat r}\nu^2\cos^2\alpha]\notag\\
&&+\gamma\left(\gamma^2\sin\alpha\frac{d\nu}{d\tau}+\nu\cos\alpha\frac{d\alpha}{d\tau}\right),\label{eq:EOM2}\\
a(U)^{\hat \varphi}&=&-\gamma^2\nu^2\sin\alpha\cos\alpha k_{\rm (Lie)}(n)^{\hat r}\notag\\
&&+\gamma\left(\gamma^2\cos\alpha\frac{d\nu}{d\tau}-\nu\sin\alpha\frac{d\alpha}{d\tau}\right). \label{eq:EOM3}
\end{eqnarray}
The geodesics equations (\ref{eq:GE}) corresponds to $\boldsymbol{a}(U)=\boldsymbol{0}$. Using Eqs. (\ref{eq:EOM1}) -- (\ref{eq:EOM2}) together with the radial component of Eq. (\ref{eq:NU}), we obtain the test particle equations of motion, described in terms of the following set of coupled ordinary differential equations of first order \cite{Defalco2020WH}
\begin{eqnarray}
\frac{\rm d\nu}{\rm dt}&=&-\frac{e^{\Phi(r)}\sin\alpha}{\gamma^2}a(n)^{\hat r},\\
\frac{\rm d\alpha}{\rm dt}&=&-\frac{e^{\Phi(r)}\cos\alpha}{\nu} [a(n)^{\hat r}+k_{\rm (Lie)}(n)^{\hat r}\nu^2],\\
\frac{\rm dr}{\rm dt}&=&e^{\Phi(r)} \nu \sin\alpha \sqrt{1-\frac{b(r)}{r}},
%
\end{eqnarray}
where we have written them in terms of the coordinate time $t$ by using the time component of Eq. (\ref{eq:NU})
\begin{equation}
\frac{\rm dt}{\rm d\tau}=\frac{\gamma}{e^{\Phi(r)}}.
\end{equation}

\subsection{The epicyclic frequencies}
\label{sec:EF}
The explicit formulas of the epicyclic frequencies $\left\{\nu_r=\Omega_r/(2\pi), \nu_\varphi=\Omega_\varphi/(2\pi)\right\}$ can  be calculated in terms of the epicyclic angular velocities $\left\{\Omega_r,\Omega_\varphi\right\}$. Defining $\boldsymbol{X}=(\nu,\alpha,r)$, the dynamical system given by Eqs. (\ref{eq:EOM1}) -- (\ref{eq:EOM3}) can be written as ${\rm d}\boldsymbol{X}/{\rm d}t=\boldsymbol{f}(\boldsymbol{X})$. We consider a stable circular orbit $\boldsymbol{X_0}=(\nu_K,0,r_0)$, where 
\begin{equation}
\nu_K\equiv\sqrt{-\frac{a(n)^{\hat r}}{k_{\rm (Lie)}(n)^{\hat r}}}=\sqrt{r\Phi'(r)},
\end{equation}
is the Keplerian angular velocity. It is easy to check that $\boldsymbol{X_0}$ is an equilibrium configuration of Eqs. (\ref{eq:EOM1}) -- (\ref{eq:EOM3}), namely $\boldsymbol{f}(\boldsymbol{X_0})=\boldsymbol{0}$. Therefore, we consider a small perturbation $\varepsilon\ll1$ around $\boldsymbol{X_0}$ given by
\begin{equation}
\nu=\nu_K+\varepsilon \nu_1, \qquad \alpha=\varepsilon \alpha_1,\qquad r=r_0+\varepsilon r_1,
\end{equation}
or also $\boldsymbol{X}=\boldsymbol{X_0}+\varepsilon\boldsymbol{X_1}$, with $\boldsymbol{X_1}=(\nu_1,\alpha_1,r_1)$. Linearizing the dynamical system, we obtain 
\begin{equation}
\frac{{\rm d}\boldsymbol{X_1}}{\rm dt}=\mathbb{A}\cdot \boldsymbol{X_1},\qquad \mathbb{A}_{ij}=\left(\frac{\partial f_i}{\partial X_j}\right)_{\boldsymbol{X}=\boldsymbol{X_0}}.
\end{equation}
Therefore, we explicitly obtain 
\begin{eqnarray}
\frac{\rm d\nu_1}{\rm dt}&=&\alpha_1 \sqrt{1-\frac{b(r_0)}{r_0}} e^{\Phi (r_0)} \Phi '(r_0) \left[r_0 \Phi '(r_0)-1\right],\label{eq:AEom1}\\
\frac{\rm d\alpha_1}{\rm dt}&=&\frac{\sqrt{1-\frac{b(r_0)}{r_0}} e^{\Phi (r_0)}}{r_0}\Bigg\{
2\nu_1\label{eq:AEom2}\\
&&\left.-r_1\frac{\left[\Phi '(r_0)+r_0 \Phi ''(r_0)\right]}{\sqrt{r_0 \Phi '(r_0)}}\right\},\notag\\
\frac{\rm dr_1}{\rm dt}&=&\alpha_1 \sqrt{1-\frac{b(r_0)}{r_0}} e^{\Phi (r_0)} \sqrt{r_0 \Phi '(r_0)}.\label{eq:AEom3}
\end{eqnarray}
To calculate the radial epicyclic angular velocity,  we must differentiate Eq. \eqref{eq:AEom2} and then use Eqs. \eqref{eq:AEom1} -- \eqref{eq:AEom3}, which implies the following harmonic oscillator equation
\begin{equation}
\frac{\rm d^2\alpha_1}{\rm dt^2}+\Omega_r^2\ \alpha_1=0,
\end{equation}
where the \emph{radial epicyclic angular velocity} $\Omega_r$ is
\begin{equation} \label{eq:Wr}
\Omega_r^2=e^{2 \Phi (r_0)}\frac{b(r_0)-r_0}{r_0}\left[2 \Phi '^2(r_0)-\frac{3\Phi '(r_0)}{r_0}-\Phi ''(r_0)\right].
\end{equation}

Since the equations of motion are rotationally invariant due to the spherical symmetry, we have that the \emph{azimuthal epicyclic angular velocity} $\Omega_\varphi$ is equal to the Keplerian angular velocity $\Omega_K$, given by \cite{DeFalco2021}
\begin{equation} \label{eq:Wf}
\Omega_K\equiv\frac{\rm d\varphi}{\rm dt}=\frac{e^{\Phi(r_0)}\nu_K}{r_0}=e^{\Phi(r_0)}\sqrt{\frac{\Phi'(r_0)}{r_0}}.
\end{equation}

\section{Applications}
\label{sec:app}
The epicyclic frequencies assume a prominent role in \emph{X-ray binaries}, which are double systems typically formed by a BH (or a neutron star) which is gravitationally bounded to its companion star. They are usually characterized by two distinctive features \cite{Lewin1997}: (1) the presence of an accretion disk formed around the compact object, which emits in all energy bands of the electromagnetic spectrum, especially with more brightness in the X-rays owed to the radiation coming from the matter inflow in the innermost regions; (2) the appearance of significant flux variabilities on long  and much shorter times-scales. The former can be appreciated on long-term light-curves and imply significant changes in the energy spectra as reported in the X-ray hardness-intensity diagrams; whereas the latter cannot be studied by investigating the light-curve, and, for this reason,  the Fourier analysis is commonly employed through power-density spectra to reveal very fast aperiodic and quasi-periodic variabilities. About last point, \emph{a feature observed in almost all kinds of accreting systems is the existence of narrow peaks with a distinct centroid frequency, well known in the literature as QPOs} (see Refs. \cite{Motta2016,Ingram2020}, for reviews).

QPOs are usually associated with accretion-related time-scales and to certain effects of strong gravity on the motion of matter around massive compact objects. Their study is extremely relevant, because they represent an astrophysical mean to explore the accretion flow around BHs in an alternative approach not accessible via energy spectra alone and can provide also indirect tests of gravity within/without GR theory \cite{Motta2016}. Although they are strong and easily measurable signals, their physical origin remains still matter of debate. However, many models have been proposed so far to explain the origin and the evolution of QPOs in X-ray binaries, which contributed thus to increase our understanding toward their observational and theoretical characteristics. 
 
An interesting aspect of all QPO models relies on the fact that they share an extensive use of the epicyclic frequencies through disparate theoretical treatments to describe the matter motions in the vicinity of BHs. Therefore, we can generally state that \emph{the observations of the epicyclic frequencies can be associated with QPO measurements}. The detection of this phenomenon involves also general relativistic light bending effects in the strongly curved BH spacetime, and polarization measurements, which allows to distinguish between the different proposed QPO models \cite{Beheshtipour2016}. The acquisition of the observational data can be performed by the actual telescopes, like Rossi X-ray Timing Explorer (RXTE) \cite{Remillard2006}, and by near-future space-missions, like LOFT (Large Observatory for X-ray Timing) \cite{Feroci2016}, eXTP (Enhanced X-ray Timing and Polarization mission) \cite{Zhang2016}, IXPE (Imaging X-ray Polarimetry Explorer) \cite{Soffitta2013}.

The aim of this digression is critical to raise the awareness on the strong observational power of the epicyclic frequencies and on the consequent great possibility to experimentally achieve the objectives proposed in this paper. This section is dedicated to the applications of the results obtained in Sec. \ref{sec:WHEPI}. We first show how to detect WHs through epicyclic frequencies (see Sec. \ref{sec:WHdet}) and then we illustrate how to reconstruct a WH solution through the observational data (see Sec. \ref{sec:WHrec}).

\subsection{Wormhole's detection: deviations from a Schwarzschild black hole}
\label{sec:WHdet}
To determine the presence of a WH, we should be able to \emph{detect metric-departures from the Schwarzschild BH geometry}. A direct and simple approach to achieve this goal can be performed by comparing the epicyclic frequencies of the Schwarzschild spacetime with those detected. Since we do not have yet observational data, in Table \ref{tab:Table1} we selected, from the literature, different WH solutions framed both in GR and in some Alternative Theories of Gravity, which can be considered straightforward extensions of GR 
\cite{Cai2015,Capozziello2009,Olmo2011,Capozziello:2011et,Capozziello2012,Clifton2012}. Below, we sketch their main features in view of WH detection. 
\begin{itemize}  
\item[$(i)$] \emph{Metric Theories:} In this class of theories, the variable describing the gravitational field is the metric tensor. A purely metric Lagrangian, linearly depending on the Ricci scalar (the Einstein-Hilbert Larangian), is necessary to have a second order dynamics. Considering non-linear combinations of curvature invariants give higher--order Lagrangians with fourth-order field equations as, e.g., the so-called $f(R)$ gravity. Adopting different forms of $f$, it is possible to address a wide range of significant phenomena at infrared scales, like: clustering of structures and accelerated expansion of the Hubble flow. The key-feature of this approach relies in solving the dark side problem through geometry. It exploits the more degrees of freedom of the gravitational field to model the constituents of dark energy and dark matter, without searching for new exotic material components, but standard perfect fluids can be notwithstanding employed. 
%
\item[$(ii)$] \emph{Metric--Affine Theories:} This class of theories concerns a  generalisation of the metric approach, because it considers metric and connection  as independent fields, allowing thus the matter to couple not only with  metric, but also with  connection. In addition, some of these theories can be   formulated relaxing the hypothesis of {\it metricity}, which practically means considering the Equivalence Principle at the foundation of gravitational interaction and then the coincidence of the causal and geodesic structures of the spacetime. An example of this formalism is represented by the Palatini formulation,  where metric and affine connections are not necessarily related through the Levi-Civita connection.
\item[$(iii)$] \emph{Teleparallel Theories:} The central features of this class of theories are: $(i)$ the Lagrangian of the gravitational field is a function $f$ of the torsion scalar $T$; $(ii)$ the Weitzenb\"{o}ck connection, is adopted instead of  the Levi-Civita connection; $(iii)$ a geodesic structure is not necessary but dynamics and kinematic are ruled by affinities. This formulation shares a deep analogy with GR, because the field equations, written in terms of $T$, can be rearranged in the same way of those expressed in terms of $R$. However, despite of the fact that Teleparallel Equivalent General Relativity (TEGR) practically coincide with GR from the point of view of the field equations,  discrepancies emerges for  the metric $f(R)$ formulation and $f(T)$ gravity. They are: (1) the $f(R)$ field equations are of fourth-order, whereas the $f(T)$ field equations remain of second-order (as it also occurs in GR); (2) the dynamical variables in the teleparallel gravity are the tetrad fields $e^\mu_\alpha$, while, in the metric formulation, this role is fulfilled by the metric tensor $g_{\mu\nu}$. This means the breaking of the Lorentz invariance and other differences which emerge in the two formulations. 
\end{itemize}
It is important to note that the procedure and calculations we have performed in Secs. \ref{sec:TLG} and \ref{sec:EF} to derive the epicyclic frequencies in generic WH spacetimes are valid not only for metric theories of gravity, but also for metric-affine and teleparallel models. Indeed, the latter two theories can be reduced to \emph{equivalent metric theories}, where the further degrees of freedom of the geometrical background are encoded in the metric potentials $\Phi(r)$ and $b(r)$ (see Refs. \cite{Capozziello:2011et,Cai2015}, for more details).


Before showing the method, we recall that the Schwarzschild metric can be defined by Eq. \eqref{eq:MTmetric} through
\begin{equation} \label{eq:SCH}
\Phi(r)=\frac{1}{2}\log\left(1-\frac{2M}{r}\right),\qquad b(r)=2M.
\end{equation}
In a Schwarzschild BH spacetime, the innermost stable circular orbit (ISCO) radius is $r_{\rm ISCO}=6M$, while the two epicyclic angular velocities are \cite{Abramowicz2005}
\begin{equation}
\Omega_r^G=\sqrt{\frac{M}{r_0^3}\left(1-\frac{6M}{r_0}\right)},\qquad \Omega_\varphi^G=\sqrt{\frac{M}{r_0^3}}.
\end{equation}

Regarding static and spherically symmetric WHs, we already calculated the epicyclic angular velocities \eqref{eq:Wr} and \eqref{eq:Wf}. The ISCO radius can be calculated through the formula (see Ref. \cite{DeFalco2021}, for  details)
\begin{equation} \label{eq:ISCO}
L^2_z[\Phi'(r)r-1]+\Phi'(r)r^3=0,    
\end{equation}
where $L_z$ is the conserved angular momentum with respect to $\boldsymbol{z}$-axis, orthogonal to the equatorial plane, along the test particle trajectory. Equation \eqref{eq:ISCO}, solved for the lowest value of $L_z$, permits to determine $r_{\rm ISCO}$. 
\begin{figure*}[t!]
\centering
\hbox{
\includegraphics[scale=0.43]{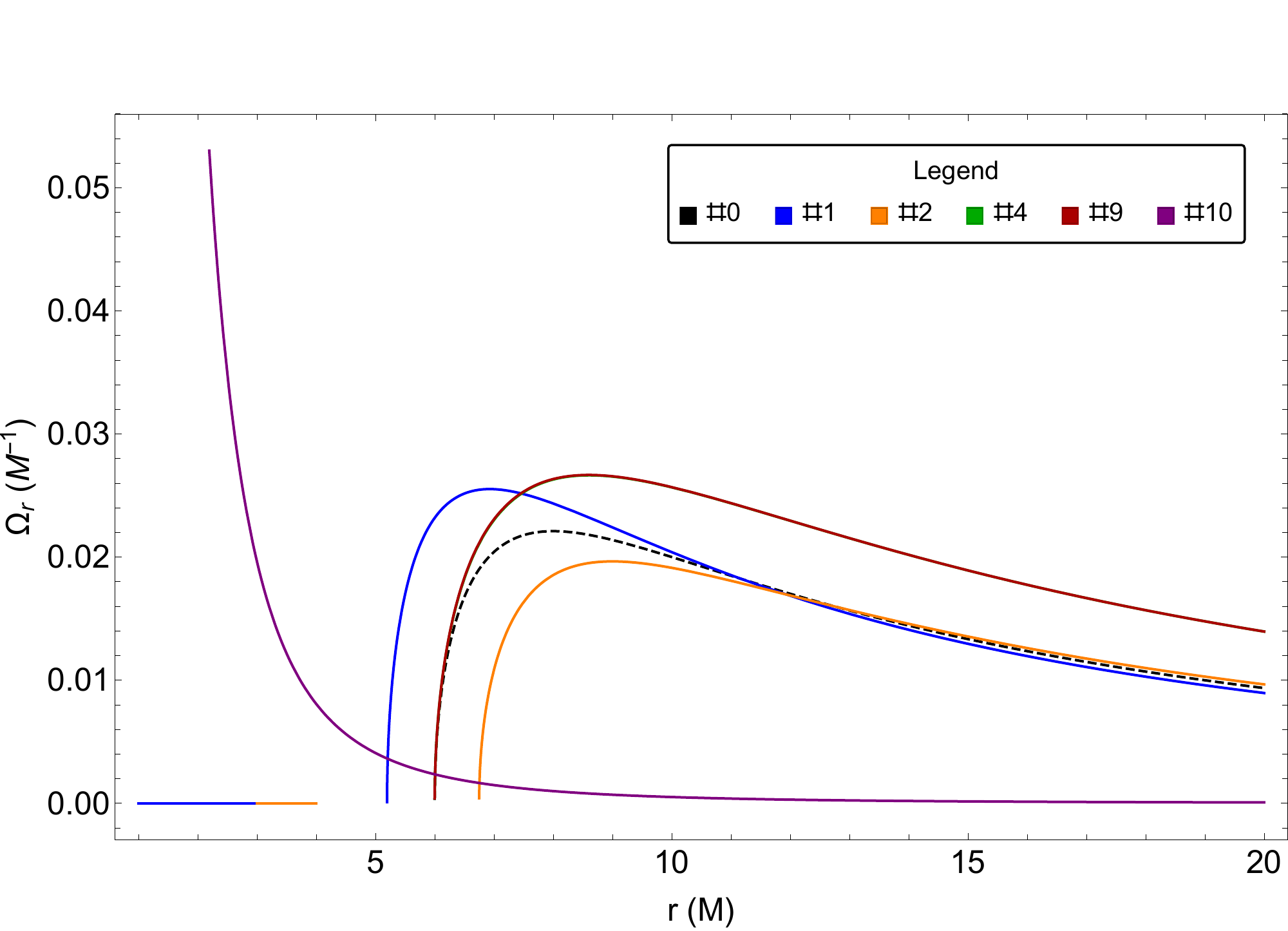}
\hspace{0.2cm}
\includegraphics[scale=0.43]{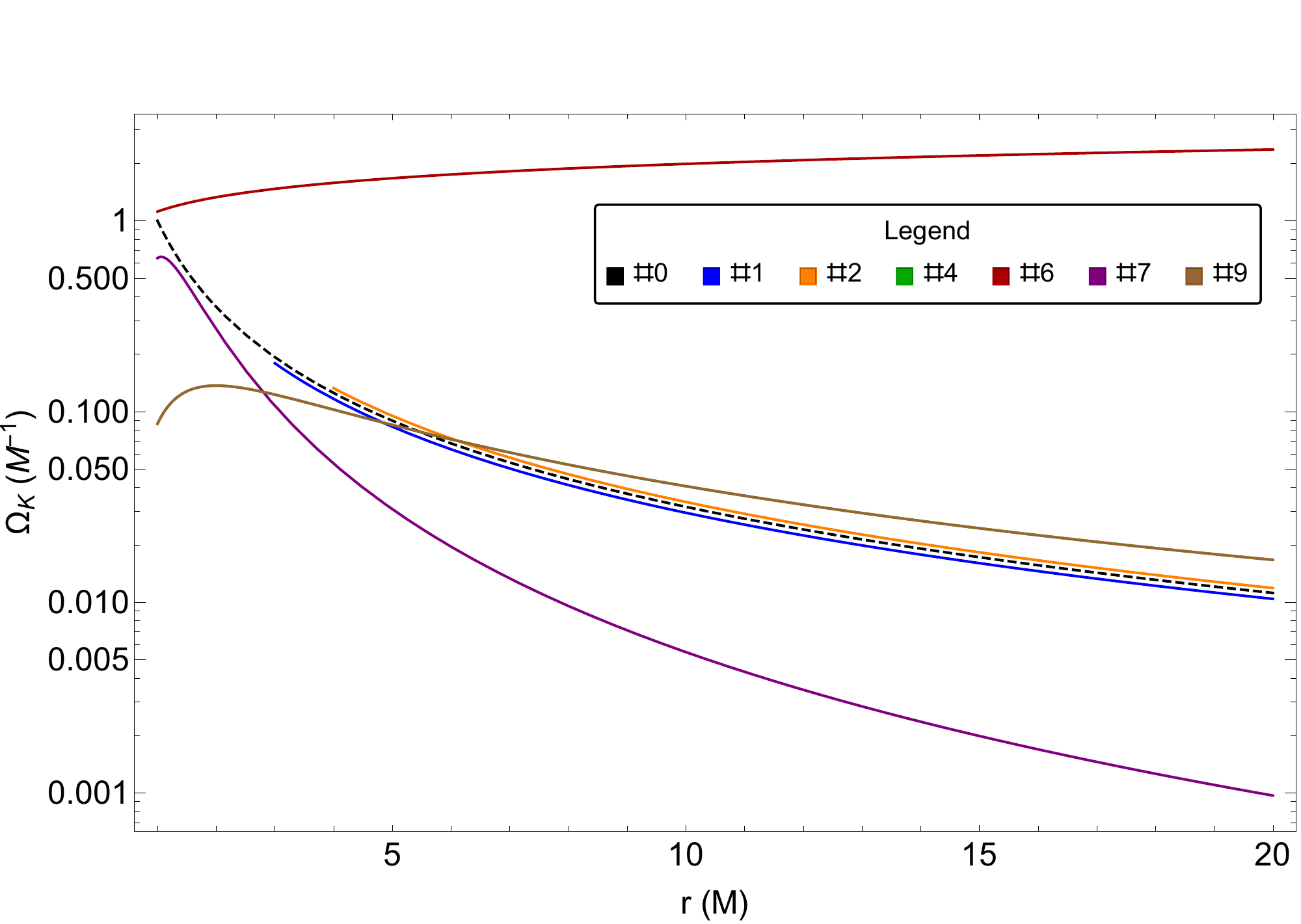}}
\caption{Plots of the radial (\emph{left panel}) and azimuthal (\emph{right panel}) epicyclic angular frequencies in terms of the radius $r$ of some WH solutions, whose numbers in the legends refer to those reported in Table \ref{tab:Table1}. The dashed black line represents the Schwarzschild BH case.}
\label{fig:Fig3}
\end{figure*}
In Fig. \ref{fig:Fig3} we show the trend of the epicyclic angular velocities of some WH solutions reported in Table \ref{tab:Table1}, namely only those that admit real values. We include also the Schwarzschild values in order to show how some WH solutions can closely emulate the BH behaviour. 

\begin{figure*}[t!]
\centering
\hbox{
\includegraphics[trim=0cm -0.3cm 0cm 0cm,scale=0.43]{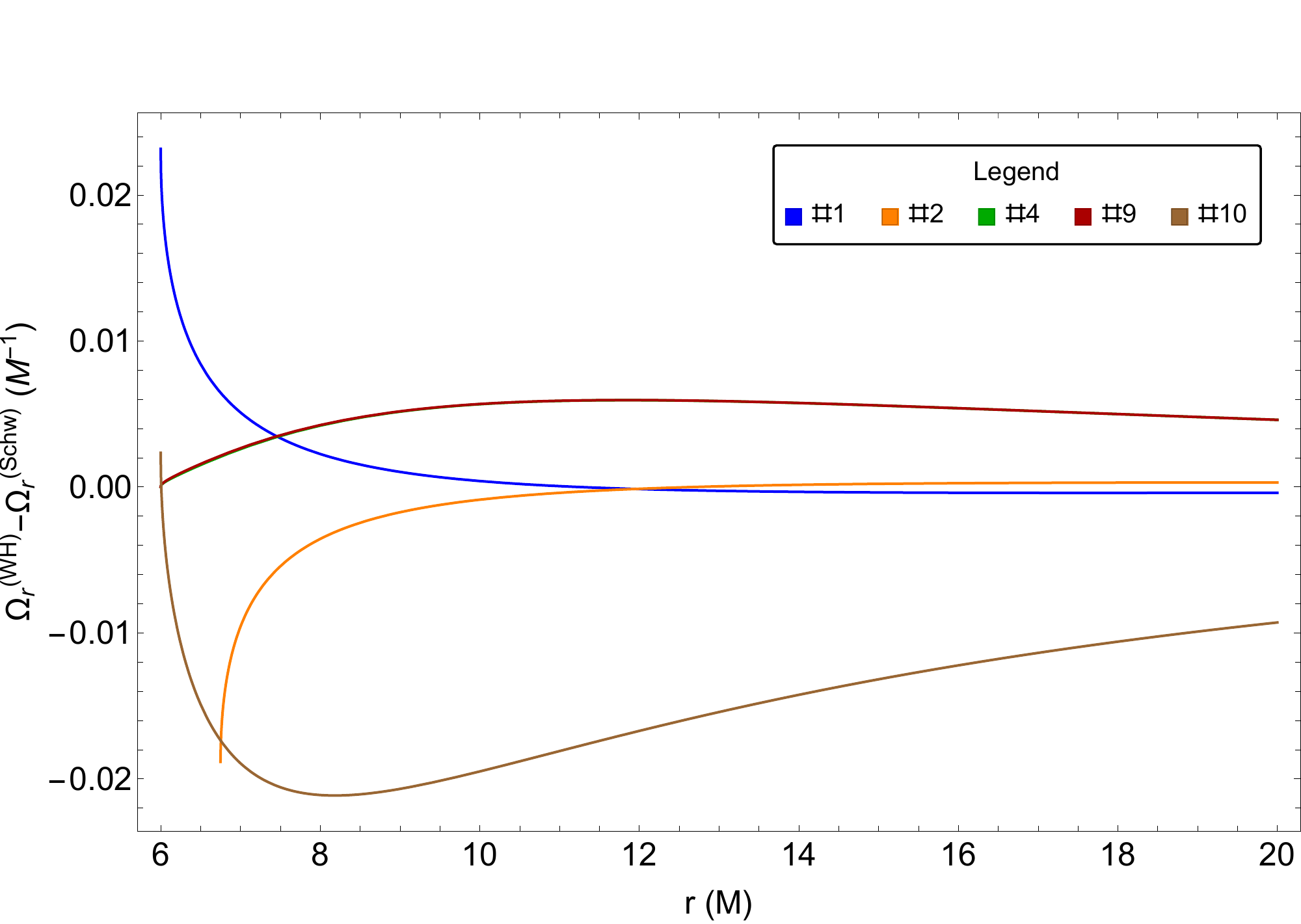}
\hspace{0.2cm}
\includegraphics[scale=0.43]{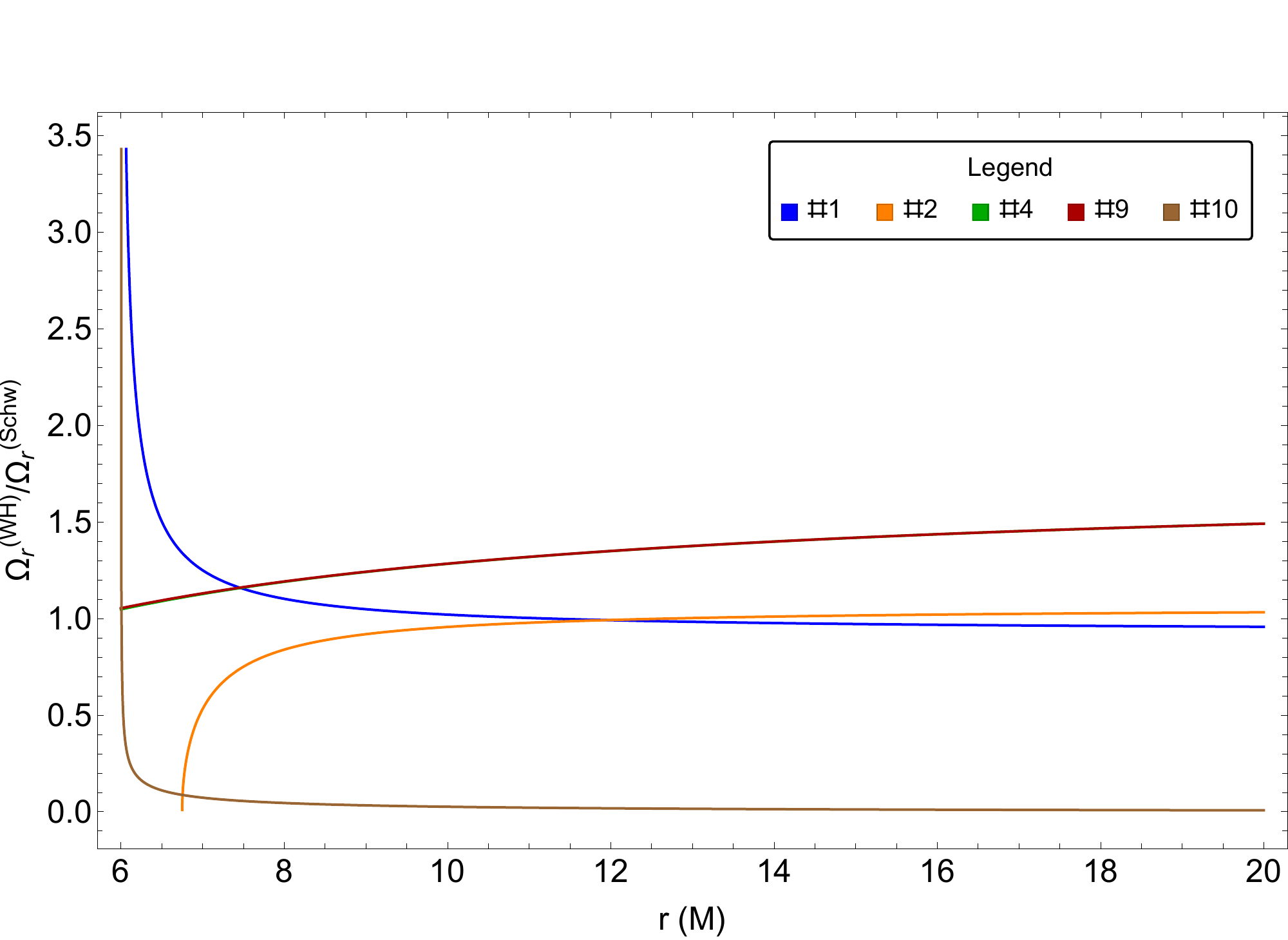}}
\vspace{0.3cm}
\hbox{
\includegraphics[scale=0.43]{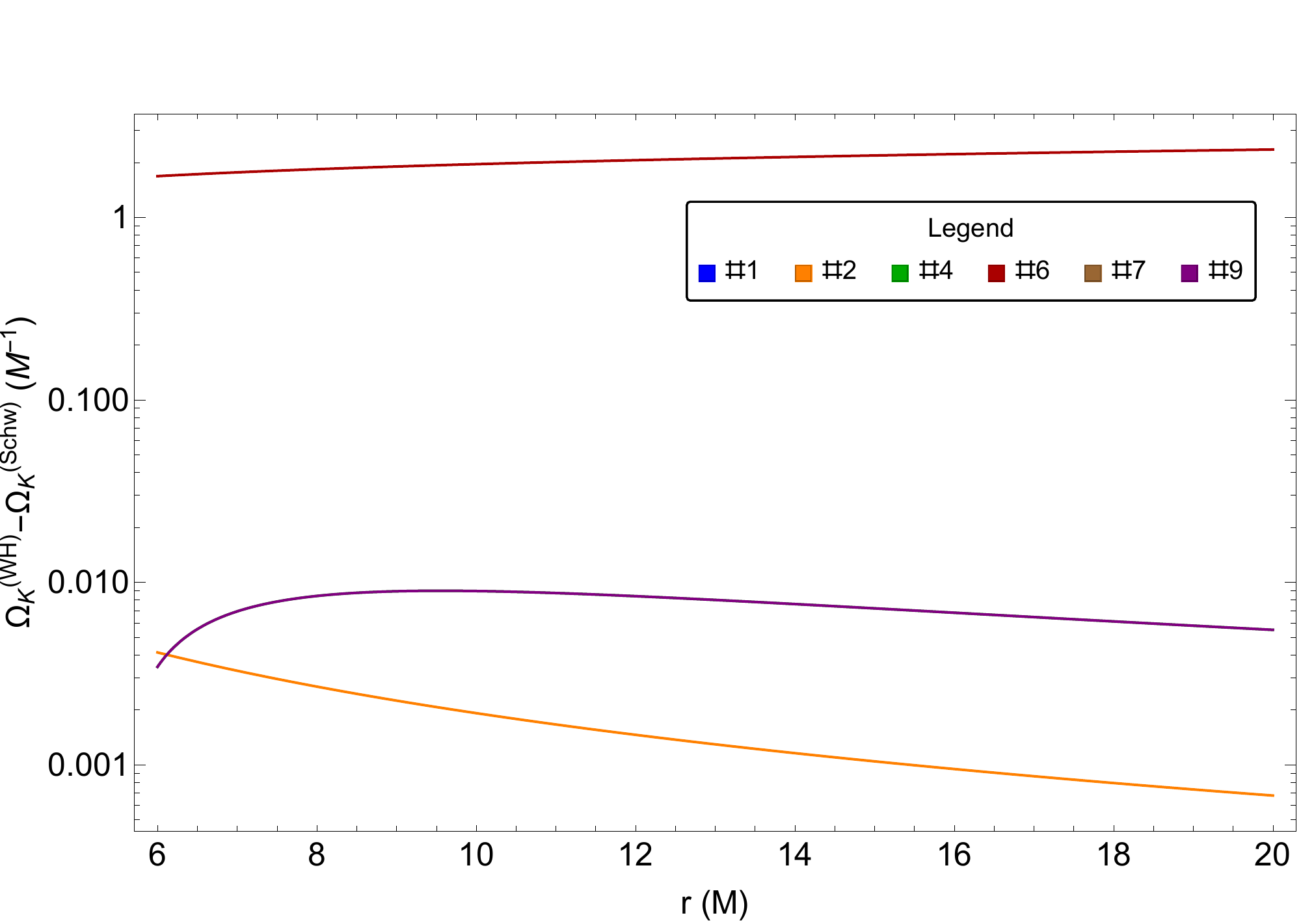}
\hspace{0.2cm}
\includegraphics[scale=0.43]{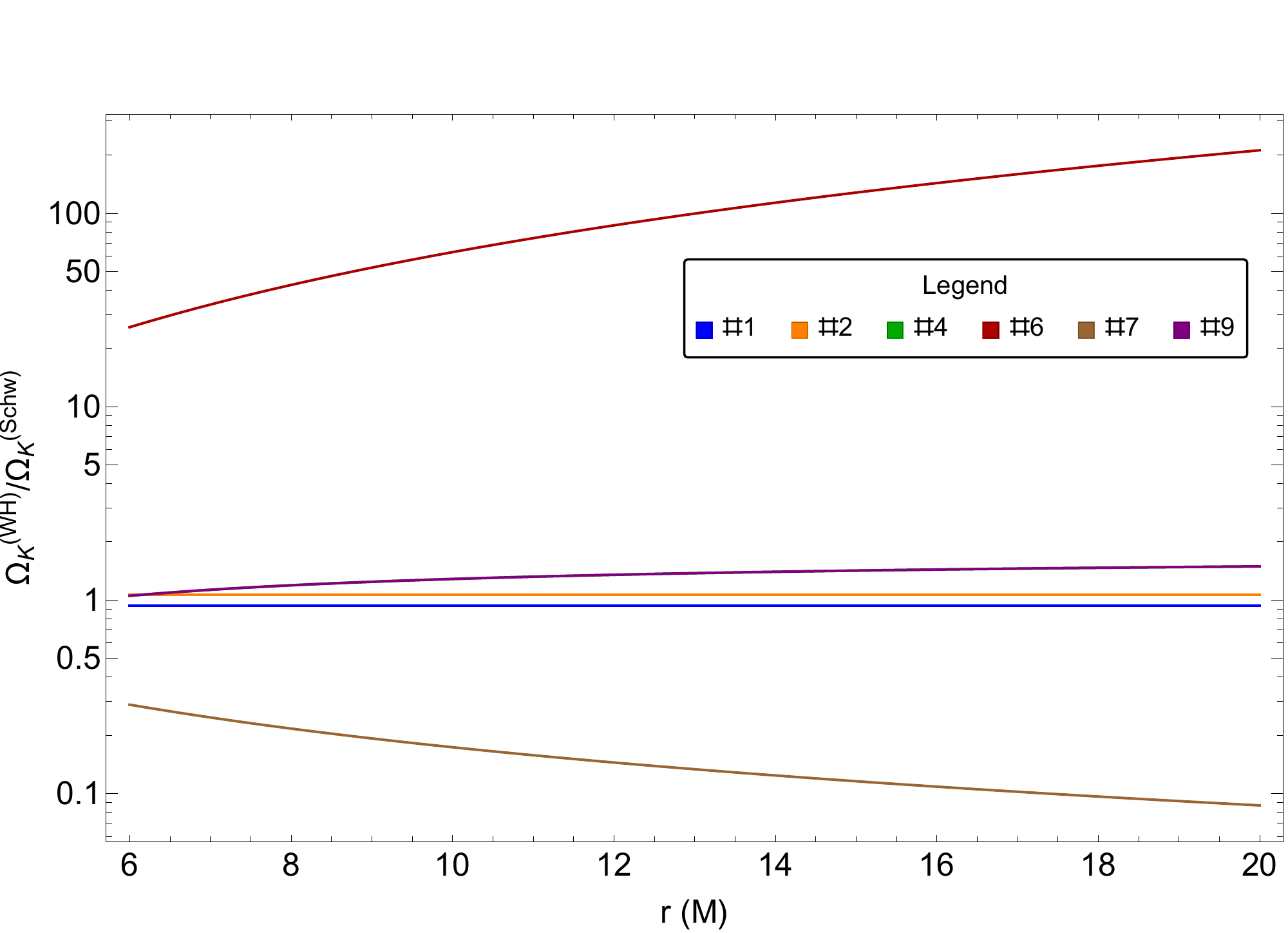}}
\caption{Plots of difference (\emph{left panels}) and ratio (\emph{right panels}) between the radial and azimuthal epicyclic angular frequencies for some WH solutions $\left\{\Omega_r^{\rm (WH)},\Omega_K^{\rm (WH)}\right\}$ (only those admitting real values, see Table \ref{tab:Table1} and Fig. \ref{fig:Fig4}), and the Schwarzschild BH values $\left\{\Omega_r^{\rm (Schw)},\Omega_K^{\rm (Schw)}\right\}$ (represented by dashed black lines).}
\label{fig:Fig4}
\end{figure*}

The practical way to disclose possible departures from BH geometries through the acquired data is by first measuring the mass of the compact object $M$ (see Refs. \cite{Falanga2015,Bambi2016}, for details), which then implies where the ISCO radius is located for a BH. As shown in Table \ref{tab:Table1} there are some WH solutions  having an ISCO radius different from $6M$ (see $\#1$ and $\#2$). These preliminary crossed observations could reveal already at the beginning whether  indications of WH existence can be present. To better analyse the selected WH solutions, we consider both the difference and the ratio of the epicyclic frequencies with respect to the Schwarzschild known profiles, see Fig. \ref{fig:Fig4}. We take into account both possibilities to quantify how much the data approach closer to the BH physics and to have an estimate of their order of magnitude. 

From Fig. \ref{fig:Fig4}, we can deduce: (1) there exists some WH geometries that does not admit either one or both epicyclic frequencies (see $\#6, \#7,\#8,\#10$), or even they admit constant zero values (see $\#3, \#5$, because they have constant redshift function); (2) there are some WH solutions, which perfectly mimic the BH trends (see $\#1,\#2,\#9$ in the lower panels of Fig. \ref{fig:Fig4}), and others get closer to the BH profiles at larger radii (see $\#7$ in the lower panels of Fig. \ref{fig:Fig4}). For the last cases, it would be important to detect the epicyclic frequencies closer to the BH ISCO, where their discrepancies are higher.

\begin{figure}[t!]
\centering
\includegraphics[scale=0.43]{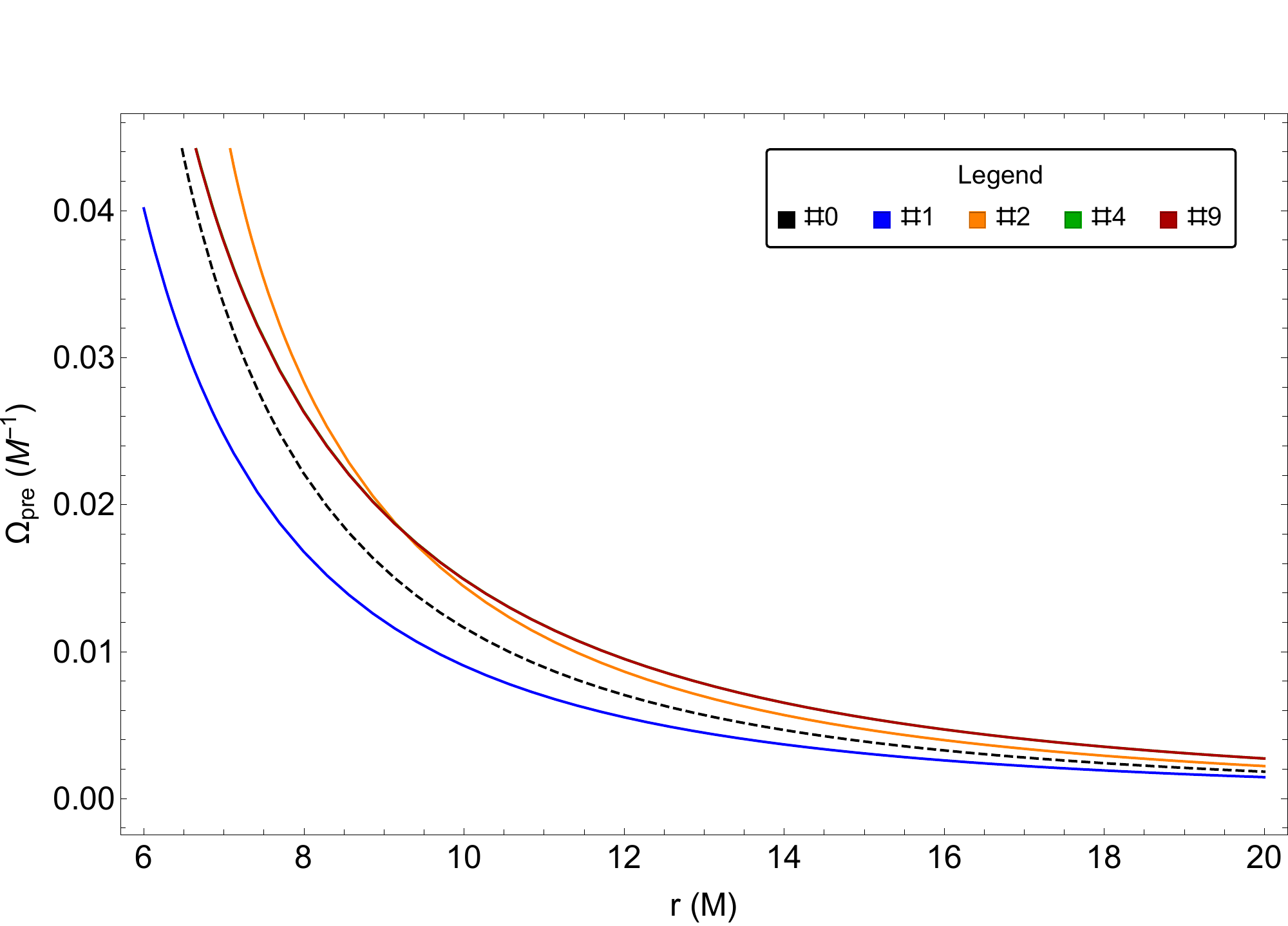}
\caption{Precession frequency $\Omega_{\rm per}$ in terms of the radius $r$ for some WH solutions admitting real values of both epicyclic angular velocities (see Table \ref{tab:Table1}). The dashed black line is the Schwarzschild BH case. The lower tight panel reports the relative errors with respect to the Schwarzschild geometry.}
\label{fig:Fig5}
\end{figure}

Another way to infer information consists in considering the precession frequency $\nu_{\rm per}=\nu_\varphi-\nu_r$ \cite{Stella1999}. In Fig. \ref{fig:Fig5}, we plot these frequencies only for those WH geometries  admitting both epicyclic frequencies. These measurements allow to set tighter constraints on the collected data, and robustly understanding the presence of a WH. 

\renewcommand{\arraystretch}{2}
\begin{table*}[t!]
\begin{center}
\caption{\label{tab:Table1} We report some examples of WH solutions in different gravity frameworks. In the first row, the parameters related to a Schwarzschild BH are reported, where $b_0$, only in this case, stays for the event horizon radius.}
\normalsize
\scalebox{0.62}{
\begin{tabular}{|@{} c@{} |@{} c @{}| @{}c @{}|@{} c @{}|@{} c @{}|@{} c @{}|@{} c @{}|@{} c @{}|@{} c@{} |} 
\ChangeRT{1pt}
{\bf Theory} & $\quad\boldsymbol{\#}\quad$ &$\quad\boldsymbol{\Phi(r)}\quad$ & $\quad\boldsymbol{b(r)}\quad$  &  $\quad\boldsymbol{b_0}\quad$  &  $\quad\boldsymbol{r_{\rm ISCO}}\quad$  & $\quad\boldsymbol{\Omega_r(r)}\quad$ & $\quad\boldsymbol{\Omega_K(r)}\quad$ &  \quad {\bf Ref.} \quad  \\
\hline
\hline
\rowcolor{lightgray} Schwarzschild &0 & $\frac{1}{2}\log\left(1-\frac{2M}{r}\right)$ & $2M$ & $2M$ & $6M$ & $\sqrt{\frac{M}{r^3}\left(1-\frac{6M}{r}\right)}$ & $\sqrt{\frac{M}{r^3}}$ & \cite{Abramowicz2005}\\
\hline
\hline
\multirow[c]{3}[3]*{GR}& 1 & $
\begin{cases}
\frac{1}{2}\log\left(1-\sqrt{\frac{1}{3}}\right) & M\le r\le3M\\
\frac{1}{2}\log\left(1-\frac{\sqrt{3}M}{r}\right) & 3M\le r
\end{cases}
$
 & $
\begin{cases}
r\sqrt{3} & M\le r\le3M\\
\sqrt{3}M & 3M\le r
\end{cases}
$
 & $M$ & $5.20M$ & $
\begin{cases}
 0. & 1\leq r\leq 3 \\
 0.93\sqrt{\frac{M(r-5.2M)}{r^4}} & 3\leq r 
\end{cases}
$
 & $
\begin{cases}
 0. & 1\leq r\leq 3 \\
 0.93\sqrt{\frac{M}{r^3}} & 3\leq r\
\end{cases}
$
 & \cite{Lemos2003}\\ \cline{2-9}
& 2 & $
\begin{cases}
\frac{1}{2}\log\frac{7}{16} & 3M\le r\le4M\\
\frac{1}{2}\log\left(1-\frac{9M}{4r}\right) & 4M\le r
\end{cases}
$
 & $
\begin{cases}
\frac{9}{16}r & 3M\le r\le4M\\
\frac{9}{4}M & 4M\le r
\end{cases}
$
 & $3M$ & $6.75M$ & $
\begin{cases}
 0. & 3\leq r\leq 4 \\
 1.06\sqrt{\frac{M\left(1-\frac{9M}{4r}\right) (r-6.75M)}{r^3 (r-2.25M)}} & 4\leq r 
\end{cases}
$
 & $
\begin{cases}
 0. & 3\leq r\leq 4 \\
 1.06\sqrt{\frac{M}{r^3}} & 4\leq r
\end{cases}
$
 & \cite{Lemos2003}\\ \cline{2-9}
& 3 & $0$ & $\frac{9M^2}{r}$ & $3M$ & NO\footnote{\qm{NO} means that does not exist the ISCO radius.} & $0$ & $0$ & \cite{Myrzakulov2016}\\ \cline{2-9}
\hline
\hline
\multirow[c]{3}[3]*{METRIC}& 4 & $-\frac{3M}{r}$ & $\frac{r}{e^{r/M-1}}$ & $M$ & $6M$ & $\sqrt{3} \sqrt{\frac{e^{-\frac{r}{M}-\frac{6M}{r}} \left(e^{\frac{r}{M}}-e\right)M (r-6M)}{r^4}}$ & $\sqrt{3} e^{-\frac{3M}{r}} \sqrt{\frac{M}{r^3}}$ & \cite{Godani2020}\\ \cline{2-9}
& 5 & $0$ & $\frac{M^3}{r^2}$ & $M$ & NO & $0$ & $0$ & \cite{Sharif2013}\\ \cline{2-9}
& 6 & $-\frac{1}{2}\log\left[\left(\frac{M}{r}\right)^{2.5}\right]$ & $r^{0.83}$ & $M$ & NO & ${}^{(*)}\footnote{The symbol ${}^{(*)}$ means that the marked functions have not real values.}$  $0.79 \sqrt{\frac{r^{0.33} \left(M^{0.17}-r^{0.17}\right)}{M^{2.5}}}$ & $\frac{1.12 r^{0.25}}{M^{1.25}}$ & \cite{Calza2018}\\ \cline{2-9}
\hline
\hline
\multirow[c]{2}[2]*{\quad METRIC-AFFINE\quad}& 7 & $-\frac{M^3}{r^3}$ & $-\frac{M^2}{r}$ & $M$ & NO & ${}^{(*)}\ \frac{\sqrt{3} e^{-\frac{M^3}{r^3}} \sqrt{-M^3(r-M)(r+M) \left(r^3+6M^3\right)}}{r^5}$ & $ \sqrt{3} e^{-\frac{M^3}{r^3}}\sqrt{\frac{M^3}{r^{5}}}$ & \cite{Capozziello2012}\\ \cline{2-9}
& 8 & $\frac{1}{2}\log[ A(r)]$\footnote{We have that $A(r)=\frac{1}{\Omega_+(r/M)}\left[1-\frac{1+0.25G(r/M)}{0.5\sqrt{\Omega_-(r/M)}r/M}\right]$, where $\Omega_\pm(z)=1\pm z^{-4}$ and $G(z)=-0.57+0.5\sqrt{z^4-1}\left[f_{3/4}(z)+f_{7/4}(z)\right]$ with $f_\lambda(z)={}_2F_1(1/2,\lambda,3/2,1-z^4)$ being the hypergeometric function. We indicate with $'=d/dr$.} & $r[1-A(r)\Omega_+^2(r/M)]$ & $M$ & NO & ${}^{(*)}\ 0.71 \sqrt{\frac{\Omega_+(r)^2 \left(A(r) \left(3. A'(r)+1. r A''(r)\right)-2. r A'(r)^2\right)}{r}}$ & ${}^{(*)}\ 0.71 A(r)^{1.5} \sqrt{\frac{A'(r)}{r A(r)}}$ & \cite{Bejarano2017}\\ \cline{2-9}
\hline
\hline
\multirow[c]{2}[2]*{\quad TELEPARALLEL\quad }& 9 & $-\frac{3M}{r}$ & $-\frac{M^3}{r^2}$ & $M$ & $6M$ & $\sqrt{3} \sqrt{\frac{e^{-\frac{6M}{r}} M(r-6M) \left(r^3+M^3\right)}{r^7}}$ & $\sqrt{3} e^{-\frac{3M}{r}}\sqrt{\frac{M}{r^{3}}}$ & \cite{Sharif2018}\\ \cline{2-9}
& 10 & $-\frac{1}{2}\log\left[\frac{1}{2}\left(1+\sqrt{1-\frac{M^2}{r^2}}\right)\right]$ & $\frac{M^2}{r}$ & $M$ & NO & $\sqrt{\frac{M^4}{r^3 \sqrt{r^2-M^2} \left[2 r \left(\sqrt{r^2-M^2}+r\right)-M^2\right]}}$ & ${}^{(*)}\ \sqrt{-\frac{M^2}{r^2 \left(\sqrt{r^2-M^2}+r\right)^2-M^2}}$ & \cite{Bohmer2012}\\ \cline{2-9}
\ChangeRT{1pt}
\end{tabular}}
\end{center}
\end{table*}

\subsection{Procedure for reconstructing the WH solution from the fit of the observational data}
\label{sec:WHrec}
In this section we propose a method to reconstruct the WH solution through the fit of the data. We first present the theoretical strategy (see Sec. \ref{sec:THEORY}) and then applies it to some simulated data (see Sec. \ref{sec:EXWH}).

\subsubsection{Theoretical strategy}
\label{sec:THEORY}
We assume that the epicyclic frequencies $\left\{\nu_r,\nu_\varphi\right\}$ (or $\left\{\Omega_r,\Omega_\varphi\right\}$) are measured in the range $r\in[r_1,r_2]$. We note that Eq. \eqref{eq:Wf} can be arranged in the following way
\begin{equation} 
\Omega_\varphi^2(r)=e^{2\Phi(r)}\frac{\Phi'(r)}{r}\quad \Rightarrow\quad \left[e^{2\Phi(r)}\right]'=2r\Omega_\varphi^2(r).
\end{equation}
Integrating both members of the above equation between $[r_1,r]$ with $r\in[r_1,r_2]$, we obtain
\begin{equation} \label{eq:Obsf}
\Phi(r)=\frac{1}{2}\log\left[e^{2\Phi(r_1)}+\int_{r_1}^r2x\Omega_\varphi^2(x) dx\right].
\end{equation}
Since the term $e^{2\Phi(r_1)}$ is unknown, it can be calculated exploiting different (and complementary) techniques, like: measuring the photon impact parameter $b_{\rm ph}(r_1)$, the photon emission angle $\alpha_E$, and the position $r_1$ at which the photon has been emitted it is possible to determine $e^{\Phi(r_1)}=r_1\sin\alpha_E/b_{\rm ph}(r_1)$,  \cite{Bisnovatyi2015,DeFalco2021}; invoking gravitational redshift effects, which permit to have $1+z=e^{\Phi(r_1)}$ \cite{Muller2010,Herrmann2018,Pumpo2021}; in the case where $r_2\gg r_1$, it would be reasonable to have $e^{2\Phi(r_1)}\approx1$, and so recast Eq. (\ref{eq:Obsf}) in a way that $e^{2\Phi(r_2)}$ appears, instead of $e^{2\Phi(r_1)}$.

On the other hand, the integral term can be exactly calculated depending on how $\Omega_\varphi$ is sampled in the region $[r_1,r_2]$. Let us assume  $N+1$ points $r_1\equiv x_0<x_1\dots<x_{N-1}<x_N\equiv r_2$, in correspondence of which we have our measured samples $\left\{\Omega_\varphi(\bar{x}_i)\right\}_{i=1,\dots,N}$, where $\bar{x}_i\in[x_{i-1},x_{i}]$ for $i=1,\dots,N$. Therefore, we arrive to the formula
\begin{equation} \label{eq:INT}
\int_{r_1}^{r_2}2x\Omega_\varphi^2(x) dx=\sum_{i=1}^N2\bar{x}_i\Omega_\varphi^2(\bar{x}_i)(x_{i}-x_{i-1}),
\end{equation}
which allows to obtain the $N$ nodes $\left\{\bar{x}_i,\Phi(\bar{x}_i)\right\}_{i=1,\dots,N}$ to be fitted. We finally reconstruct the explicit expression of $\Phi(r)$, and we can also calculate $\Phi'(r)$ and $\Phi''(r)$. 

Now instead, using Eq. \eqref{eq:Wr} we can determine the shape function $b(r)$ in $[r_1,r_2]$ through the formula 
\begin{equation} \label{eq:bOBS}
b(r)=\left[\frac{\Omega_r^2(r)}{e^{2 \Phi (r)}\left[2 \Phi '^2(r)-\frac{3\Phi '(r)}{r}-\Phi ''(r)\right]}+1\right]r.
\end{equation}
We then discretize the interval $[r_1,r_2]$ and use the observed samples $\left\{\Omega_r(\bar{x}_i)\right\}_{i=1,\dots,N}$. In this way, we straightforwardly obtain the nodes $\left\{\bar{x}_i,b(\bar{x}_i)\right\}_{i=1,\dots,N}$, which in turn can be fitted to reconstruct also $b(r)$.

\subsubsection{A test-example}
\label{sec:EXWH}
This section aims at exhibiting a practical example to further clarify how the above outlined theoretical procedure works. We provide some simulated data to compensate for the actual absence of real data on WHs.

Let us assume to measure the radial $\nu_r$ and azimuthal $\nu_\varphi$ epicyclic frequencies related to a WH of mass $M=10^5M_\odot$ in $N=10$ points $\left\{\bar{x}_i\right\}_{i=1,\dots,10}$, included in the radial interval $[r_1=7M,r_2=10M]$, see Table \ref{tab:Table2}. Let us then calculate the related epicyclic angular velocities $\left\{\Omega_r,\Omega_\varphi\right\}$, and, since we know the WH mass $M$, we can further convert them into geometrical units, see Table \ref{tab:Table2}. Furthermore, we divide uniformly the interval $[r_1,r_2]$ in bins of amplitude $\Delta r=(r_2-r_1)/10=0.3M$. This means that there exists $N+1=11$ points such that $r_1=x_0<x_1<\dots<x_9<x_{10}=r_2$ and $x_i-x_{i-1}=\Delta r$ for all $i=1,\dots,10$.
\renewcommand{\arraystretch}{2}
\begin{table}[t!]
\begin{center}
\caption{\label{tab:Table2} We report the $N=10$ points $\bar{x}_i\in[r_1=7M,r_2=10M]$ in correspondence of which we have the sampled values of the radial and azimuthal epicyclic frequencies $\left\{\nu_r,\nu_\varphi\right\}$, and angular velocities $\left\{\Omega_r,\Omega_\varphi\right\}$ (both in dimensional and geometrical units). We stress that these are not real data.}
\normalsize
\scalebox{0.75}{
\begin{tabular}{| c || c | c | c || c | c | c |} 
\ChangeRT{1pt}
$\boldsymbol{\bar{x_i}}$ &$\boldsymbol{\nu_r}$ & $\boldsymbol{\Omega_r}$  &  $\boldsymbol{\Omega_r}$  &  $\boldsymbol{\nu_\varphi}$  & $\boldsymbol{\Omega_\varphi}$ & $\boldsymbol{\Omega_\varphi}$ \\
$\boldsymbol{({\rm M})}$ &$\boldsymbol{(10^{-3}{\rm Hz})}$ & $\boldsymbol{({\rm rad}/s)}$  &  $\boldsymbol{(10^{-3}{\rm M^{-1}})}$  & $\boldsymbol{(10^{-3}{\rm Hz})}$ & $\boldsymbol{({\rm rad}/s)}$  &  $\boldsymbol{({\rm M^{-1}})}$ \\
\hline
\hline
7.17 & 7.07 & 0.044 & 2.40 & 1.75 & 0.110 & 0.059 \\ 
\hline
7.52 & 7.46 & 0.047 & 2.53 & 1.66 & 0.104 & 0.056 \\
\hline
7.84 & 7.68 & 0.048 & 2.61 & 1.59 & 0.100 & 0.054 \\ 
\hline
8.04 & 7.76 & 0.049 & 2.63 & 1.54 & 0.097 & 0.052 \\ 
\hline
8.44 & 7.84 & 0.049 & 2.66 & 1.46 & 0.092 & 0.049 \\ 
\hline
8.55 & 7.85 & 0.049 & 2.66 & 1.44 & 0.090 & 0.049 \\
\hline
8.91 & 7.83 & 0.049 & 2.66 & 1.37 & 0.086 & 0.047 \\ 
\hline 
9.11 & 7.80 & 0.049 & 2.65 & 1.34 & 0.084 & 0.045 \\ 
\hline
9.57 & 7.70 & 0.048 & 2.61& 1.26 & 0.079 & 0.043 \\ 
\hline
9.82 & 7.62 & 0.048 & 2.59 & 1.22 & 0.077 & 0.042 \\
\ChangeRT{1pt}
\end{tabular}}
\end{center}
\end{table}

We need first to determine the redshift function by exploiting Eq. \eqref{eq:Obsf}. Let us assume, we are able to provide the estimation $e^{\Phi(r_1)}=1.15$ through some observational technique. Calculating the integral \eqref{eq:INT} through the data of Table \ref{tab:Table2}, we are finally able to obtain the nodes for the redshift function $\left\{\bar{x}_i,\Phi(\bar{x}_i)\right\}_{i=1,\dots,10}$. 
\begin{figure}[h!]
\centering
\includegraphics[scale=0.4]{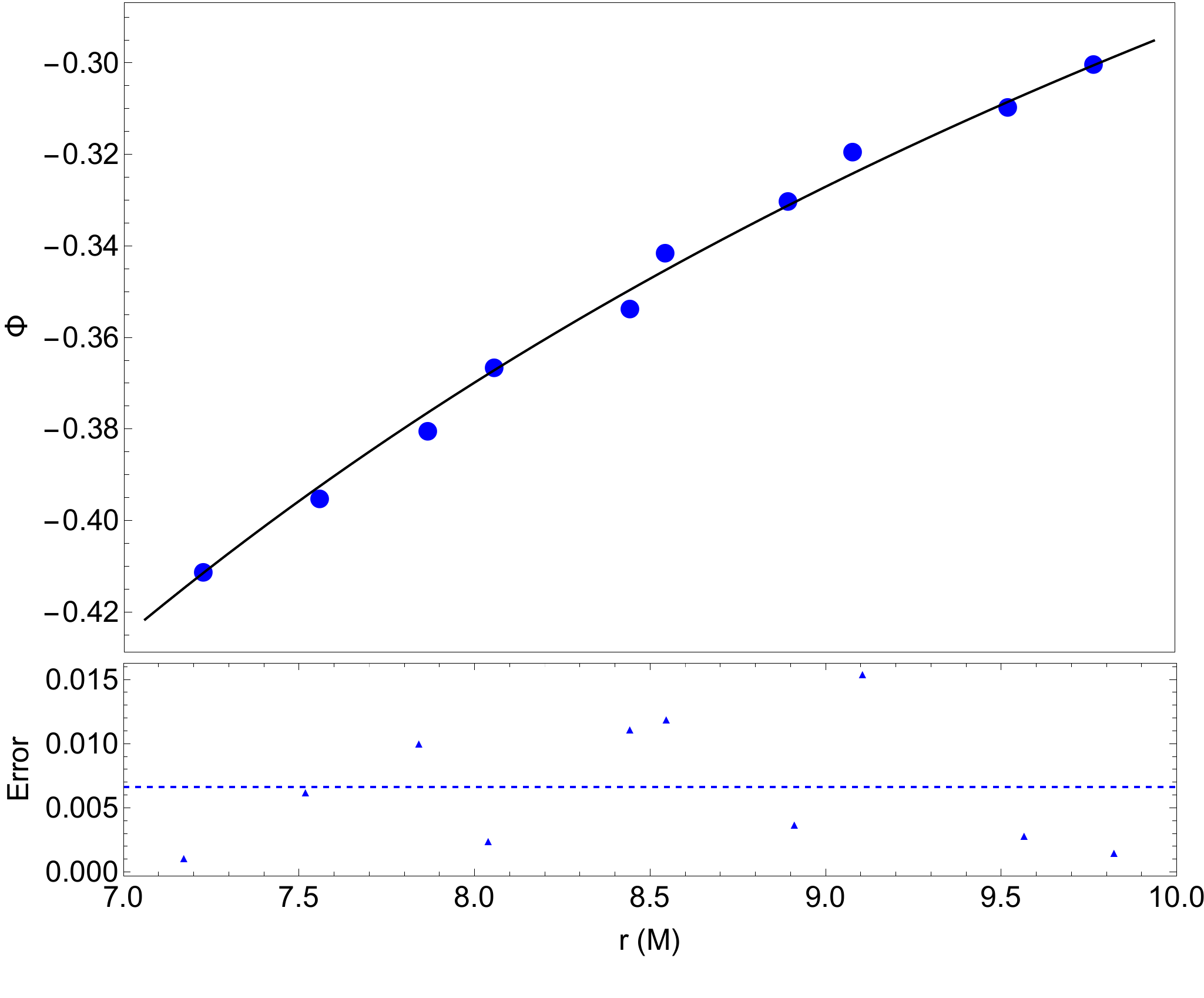}
\caption{\emph{Upper panel:} the blue points represent the nodes $\left\{\bar{x}_i,\Phi(\bar{x}_i)\right\}_{i=1,\dots,10}$, and the black line is the best fit $\Phi(r)=-2.95168/r$. \emph{Lower panel:} there are the relative fit-errors and the dashed blue line is set at the mean error $0.007$.}
\label{fig:Fig1}
\end{figure}
In Fig. \ref{fig:Fig1},  the nodes and the best fit function are represented (calculations are performed in the \texttt{Mathematica 12} environment)
\begin{equation} \label{eq:fitPhi}
\Phi(r)=-\frac{2.95168}{r},
\end{equation}
together with the \emph{relative fit-errors}\footnote{Denoted with $\left\{\Phi_i\right\}_{i=1,\dots,10}$ the real-measured values, and with $\left\{\tilde{\Phi}_i\right\}_{i=1,\dots,10}$ those values obtained through the fitting function \eqref{eq:fitPhi} evaluated at $\left\{\bar{x}_i\right\}_{i=1,\dots,10}$, then the relative fit-errors are calculated as follows $\left\{\vert\Phi_i-\tilde{\Phi}_i\vert/\vert\Phi_i\vert\right\}_{i=1,\dots,10}$.}, whose minimum, mean, and maximum values are $0.001,
0.007, 0.015$, respectively, attesting thus the good agreement of the fit. 

\begin{figure}[h!]
\centering
\includegraphics[scale=0.4]{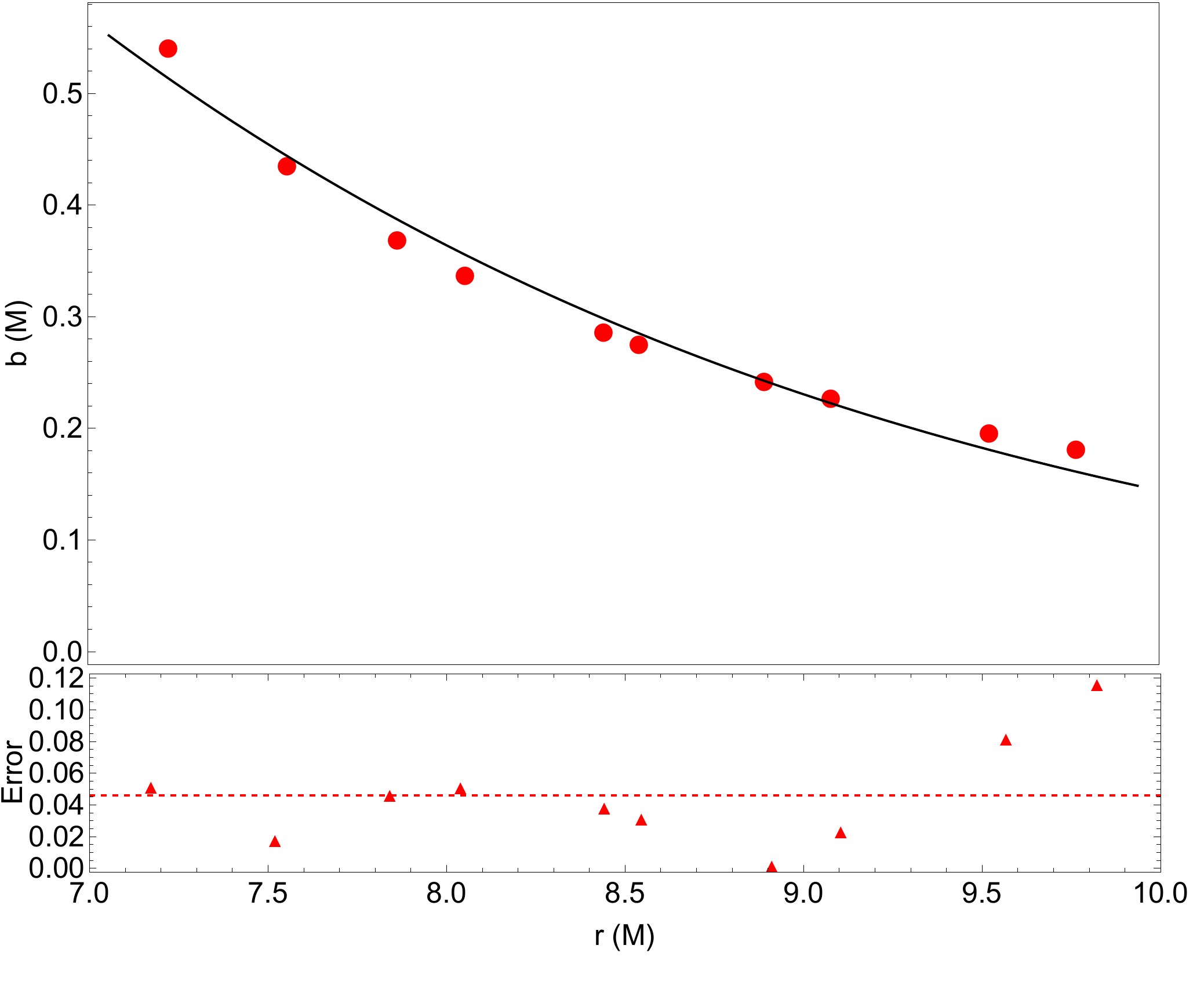}
\caption{\emph{Upper panel:} the red points represent the nodes $\left\{\bar{x}_i,b(\bar{x}_i)\right\}_{i=1,\dots,10}$, and the black line is the best fit $b(r)=r e^{1.355-0.557r}$. \emph{Lower panel:} there are the relative fit-errors and the dashed red line is set at the mean error $0.0460$.}
\label{fig:Fig2}
\end{figure}

Once we have the fitted $\Phi(r)$, we can analytically calculate $\Phi'(r),\Phi''(r)$, which permits to determine the shape function through Eq. \eqref{eq:bOBS}. Using also the values reported in Table \ref{tab:Table2}, we can calculate the nodes $\left\{\bar{x}_i,b(\bar{x}_i)\right\}_{i=1,\dots,10}$. In Fig. \ref{fig:Fig2}, nodes are fitted with the following function
\begin{equation} \label{eq:fitb}
b(r)=r e^{1.355-0.557r},
\end{equation}
whose minimum, mean, and maximum relative fit-error values are $0.0018,
0.0460, 0.1164$, respectively, confirming thus again an excellent agreement of the fit. 

\section{Discussion and conclusions}
\label{sec:end}
In this paper, we considered static and spherically symmetric WH geometries described by the Morris-Thorne-like metric \eqref{eq:MTmetric}, defined in terms of the redshift and shape functions. This theory-independent formalism provides a general approach to investigate several possible WH solutions within different gravity frameworks. We explore the inverse problem, namely the reconstruction of the WH solution through the fit of possible observational data.  

In our approach, we used the observer splitting formalism (see Sec. \ref{sec:WHEPI}), which is very useful to disentangle among gravitational and inertial effects, and it also permits to have a direct connection with the classical description, assigning therefore a precise physical meaning to the quantities theoretically manipulated, see Eqs. \eqref{eq:acc} and \eqref{eq:klie}. There is a direct way to obtain the epicyclic frequencies through the formulas
\cite{Chakraborty2017,Deligianni2021}:
\begin{eqnarray}
\Omega_\varphi&=&\sqrt{\frac{\partial_rg_{tt}}{\partial_rg_{\varphi\varphi}}},\label{eq:SWf}\\
\Omega_r&=&\frac{1}{2g_{rr}}\left[g_{tt}^2\partial_r^2g^{tt}+(\Omega_\varphi g_{\varphi\varphi})^2\partial_r^2g^{\varphi\varphi}\right],\label{eq:SWr}
\end{eqnarray}
which can be checked that are equivalent to Eqs. \eqref{eq:Wf} and \eqref{eq:Wr}, respectively. In addition, Equs. \eqref{eq:SWf} and \eqref{eq:SWr} can be very useful for numerical implementations. 

We have stressed the observational aspect of the epicyclic frequencies, since they are extensively used as a fundamental ingredient for the development of QPO models, which are easy to be detected and are a common feature in several X-ray binaries. The study of QPOs requires theoretically a general-relativistic ray-tracing code to inquire their X-ray timing spectroscopy and polarization properties, and experimentally simultaneous observations through first-generation X-ray polarimeters and LOFT-type missions. After having derived the epicyclic frequencies \eqref{eq:Wr} and \eqref{eq:Wf}, which include a combination of the redshift, together with its derivatives, and shape functions, see Sec. \ref{sec:WH}, we then applied our formulas for achieving two goals: (1) detecting the presence of a WH, distinguishing it from a BH (see Sec. \ref{sec:WHdet}); (2) exhibiting a procedure for reconstructing a WH solution through the fit of  observational data (see Sec. \ref{sec:WHrec}).

The first point is timely and essential, because it provides a further astrophysical strategy to reveal the observational existence of a WH. The method is very simple, because it relies on comparing the observed data with the BH information in order to see whether there are some relevant discrepancies. This would mean that metric-changes may occur and a WH may exist. There are some WH solutions, which closely mimic the BH observational proprieties, therefore this procedure alone is not enough sometimes and it must be complemented with other approaches presented in the literature to extract more information and tighter constraint on the theoretical models.

Once there would be available data on WHs, we should be able to reconstruct the WH solution. Specifically, exploiting Eqs. \eqref{eq:Obsf} and \eqref{eq:bOBS}, it would be possible to  reconstruct the redshift and shape functions. We provide also a test-example based on some simulated data, see Sec. \ref{sec:EXWH}. This section has only the aim to better clarify how the outlined procedure works practically with the data. Therefore, the following remarks are in order: $(i)$ the data in Table \ref{tab:Table2} may be in principle not observable; $(ii)$ the data in Table \ref{tab:Table2} are listed without detection errors (depending mainly on the instrument sensitivity used to perform the measurements), so they can be interpreted as the mean values of the detection; $(iii)$ the fit of the data and the related fit-errors can be performed with more advanced statistical methods (see e.g., \cite{Chattopadhyay2014}); $(iv)$ the sampled epicyclic frequencies and radial extent $[r_1,r_2]$ may be different from those observed.

This paper is part of a series of works aiming at providing both  observational evidences of  WH existence and different techniques to reconstruct them through potential future observational data. This model-independent approach allows not only to determine the WH solutions, but also to provide indirect observational tests of gravity within GR theory or towards Alternative Theories of Gravity. In addition, all these procedures can be adapted and extended also to study other classes of compact objects different from WHs. As near-future projects we aim at complementing this approach with other astrophysical techniques. In particular, this work can be also extended and improved for axially symmetric WHs.

\section*{Acknowledgements}
V.D.F. thanks Gruppo Nazionale di Fisica Matematica of Istituto Nazionale di Alta Matematica for the support. V.D.F., M.D.L., and S.C. acknowledge the support of INFN {\it sez. di Napoli}, {\it iniziative specifiche} TEONGRAV, QGSKY, and MOONLIGHT2.

\bibliography{references}

\end{document}